\documentclass{PoS}

\usepackage{epsfig}
\usepackage{graphics,color}
\usepackage{cite}

\newcommand{\be}{\begin{equation}}
\newcommand{\ee}{\end{equation}}

\newcommand{\bea}{\begin{eqnarray}}
\newcommand{\eea}{\end{eqnarray}}
\newcommand{\bean}{\begin{eqnarray*}}
\newcommand{\eean}{\end{eqnarray*}}

\newcommand{\om} {\omega}

\newcommand{\pv}{{\mathbf p}}

\title{P wave bottomonium spectral functions in the QGP from lattice NRQCD}

\ShortTitle{Bottomonium spectral functions in the QGP}

\author{\speaker{Gert Aarts} \\
      	Department of Physics, College of Science, Swansea University, Swansea, United Kingdom\\
        	E-mail: \email{g.aarts@swan.ac.uk}
        	}

\author{Chris Allton\\
      	Department of Physics, College of Science, Swansea University, Swansea, United Kingdom\\
      	E-mail: \email{c.allton@swan.ac.uk}
      	}

\author{Seyong Kim\\
	Department of Physics, Sejong University, Seoul 143-747, Korea\\
	E-mail: \email{skim@sejong.ac.kr}
	}
	
\author{Maria-Paola Lombardo\\
	INFN-Laboratori Nazionali di Frascati, I-00044, Frascati (RM) Italy\\
	E-mail: \email{Mariapaola.Lombardo@lnf.infn.it}
	}
	
\author{Sinead M.~Ryan\\
	School of Mathematics, Trinity College, Dublin 2, Ireland\\
	E-mail: \email{ryan@maths.tcd.ie}
	}
	
\author{Jon-Ivar Skullerud\\
	Department of Mathematical Physics, National University of Ireland Maynooth, Maynooth, County Kildare, Ireland\\
	E-mail: \email{jonivar@thphys.nuim.ie}
	}

\abstract{We present an overview of bottomonium spectral functions in the quark-gluon plasma, obtained by the {\sc fastsum} collaboration, using lattice QCD simulations with two light quark flavours on anisotropic lattices. The bottom quark is treated nonrelativistically. 
While we find that the $S$ wave ground states survive up to the highest temperature we consider, we have strong indications that $P$ wave states melt immediately above $T_c$.
   }

\FullConference{31st International Symposium on Lattice Field Theory LATTICE 2013\\
                 July 29 -- August 3, 2013\\
                 Mainz, Germany}

\begin{document}

\section{Introduction}

Quarkonia -- bound states of a heavy quark and anti-quark -- do not dissolve  immediately  in the quark-gluon plasma (QGP) \cite{Matsui:1986dk}.  
Instead it is expected that tightly bound states can survive to quite high temperatures, while broader loosely bound states decay at a lower temperature, possibly at temperatures just above $T_c$. 
By considering several ground and excited states and different heavy quark flavours  (charmonium, bottomonium), it is possible that a rich melting pattern emerges, covering temperatures from the deconfinement transition temperature $T_c$ up to four or five times $T_c$. 

 In order to turn this notion into a predictive thermometer for the QGP created at RHIC and the LHC using relativistic heavy-ion collisions, a proper theoretical understanding with control over uncertainties is required \cite{Mocsy:2013syh,Rothkopf:2013ria}.
 For bottomonium, due to the nonrelativistic nature of the $b$ quark, this might be within reach. Using the fact that the $b$ quark mass and the temperature scales are well separated, $m_b\sim 4.5$ GeV while $5T_c\lesssim$ 0.8 GeV, the heavy-quark scale can be reliably integrated out to arrive at nonrelativistic QCD (NRQCD) as an effective field theory (EFT), just as at zero temperature.
Further EFTs can be constructed by integrating out lower-energy scales, such as $m_bv$ and  $m_bv^2$, where $v$ is the heavy quark velocity in the centre-of-mass frame, and $T$ and $m_D\sim\sqrt{\alpha_s}T$, which have a thermal origin \cite{Ghiglieri:2013iya,Burnier:2007qm,Brambilla:2010vq}.

In the past few years, we have studied NRQCD  on the lattice at finite temperature for the bottomonium system and obtained a number of results in $S$ wave channels  ($\Upsilon, \eta_b$) and $P$ wave channels ($\chi_b, h_b$) \cite{Aarts:2010ek,Aarts:2011sm,Aarts:2012ka,Aarts:2013}.
 In this contribution we compare those, using the $\Upsilon$ and $\chi_{b1}$ channels as example (results in the other channels are very similar).
Our main finding is that the $S$ wave ground state appears to survive up to at least $2.09T_c$, the highest temperature we considered, while excited states are suppressed. In contrast, in $P$ wave channels we find strong indications for an immediate melting of the $P$ wave ground state, at temperatures just above $T_c$.

\section{$N_f=2$ quark-gluon plasma and correlators}

\begin{table}[b]
\begin{center}
\vspace*{0.2cm}
\begin{tabular}{| l | rrrrrrr | }
\hline
$N_\tau$ 		& 80 & 32 & 28 & 24 & 20 & 18 & 16 \\
$T$(MeV) 	& 90 & 230 & 263 & 306 & 368 & 408 & 458 \\
$T/T_c$ 		& 0.42 & 1.05 & 1.20 & 1.40 & 1.68 & 1.86 & 2.09 \\ 
$N_{\rm cfg}$   & 250 & 1000 & 1000 & 500 & 1000 & 1000 & 1000 \\
\hline
\end{tabular}
\vspace*{0.2cm}
\caption{Lattice parameters: the lattice size is $N_s^3\times
  N_\tau$ with $N_s=12$, lattice spacing $a_s\simeq 0.162$ fm, $a_\tau^{-1} =7.35(3)$ GeV, and anisotropy $a_s/a_\tau=6$ \cite{Aarts:2011sm}.  
}
\label{tab:lattice}
\end{center}
\end{table}

We start with a brief overview of lattice details \cite{Aarts:2011sm}. 
The lattices in the $N_f=2$ study are highly anisotropic, with a ratio of spatial to temporal lattice spacing of $\xi\equiv a_s/a_\tau=6$. The lattice spacing is fixed, with  $a_s\simeq 0.162$ fm. The light quarks are Wilson-like, and $m_\pi/m_\rho\simeq 0.54$. The lattice with the lowest temperature, $T/T_c=0.42$, has $N_\tau=80$ sites in the time direction, while for the one with highest temperature, $T/T_c=2.09$, $N_\tau=16$. There are 1000 configurations at most temperatures, such that correlators can be determined to very high precision.  Full details can be found in Table~\ref{tab:lattice}. 
For the $b$ quark, we use the standard ${\cal O}(v^4)$ NRQCD formulation on the lattice \cite{Davies:1994mp}. 

\begin{figure}[t]
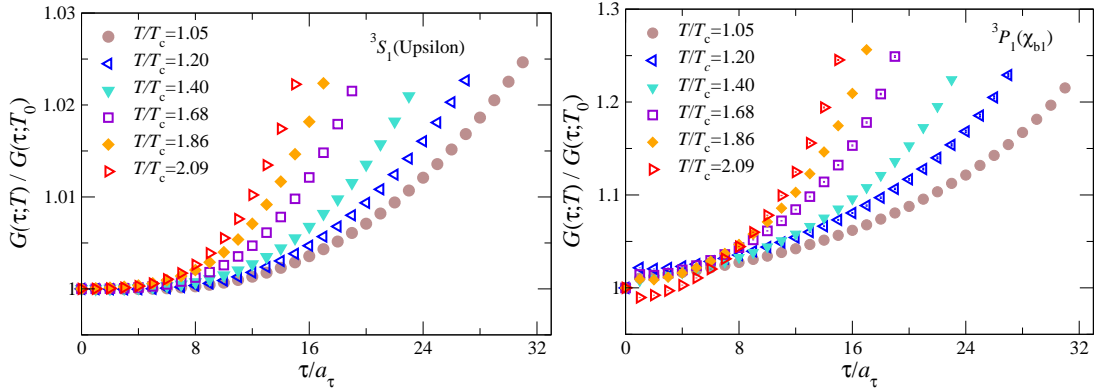

\begin{center}
\epsfig{figure=upsilon_C80_GA-v2.eps,width=0.48\textwidth}
\epsfig{figure=chi_b1_C80-GA.eps,width=0.47\textwidth}
\end{center}
 \caption{Ratio of the correlator $G(\tau;T)$ at a given temperature $T$ with the one at the lowest temperature, $T_0=0.42T_c$, in the $\Upsilon$ channel ($S$ wave) \cite{Aarts:2011sm} (left) and in the  $\chi_{b1}$ channel ($P$ wave) \cite{Aarts:2013} (right).
}
 \label{fig:1}
\end{figure}

We first discuss euclidean correlators. In Fig.\ \ref{fig:1} we show the correlators at nonzero temperature, normalised to the one at the lowest temperature, in the $\Upsilon$ channel (left) and the $\chi_{b1}$ channel (right).
 Note that we consider only zero momentum and hence do not indicate the $\pv$ dependence; a study at nonzero momentum can be found in Ref.\ \cite{Aarts:2012ka}.
 Correlators in NRQCD are not periodic. This can e.g.\ be seen from the spectral relation
\be
\label{eq:1}
G(\tau) = \int_{\om_{\rm min}}^{\om_{\rm max}} \frac{d\om}{2\pi}\, K(\tau,\om)  \rho(\om), 
\quad\quad\quad K(\tau,\om) = e^{-\om\tau},
\ee
in which the kernel $K(\tau,\om)$ is a temperature-independent exponential,  both at zero and nonzero temperature. Technically this is due to the fact that in NRQCD  quark propagators are obtained by solving an initial-value problem, rather than a boundary-value problem, as is the
case for relativistic quarks, with the kernel $K(\tau,\om)=\cosh[\om(\tau-1/2T)]/\sinh(\om/2T)$.
Physically it implies there are no thermal $b$ quarks present in the plasma.
 Hence in NRQCD there is no {\em kinematical} temperature dependence due to boundary conditions. Consequently the entire euclidean time interval, $0<\tau<1/T$, can be used. 
Nevertheless, temperature dependence is present in the correlators as a {\em dynamical} effect, since the heavy quarks propagate through a heatbath of gluons and light quarks at different temperatures.
In Fig.\ \ref{fig:1} we observe a significant difference in this temperature dependence between $S$ and $P$ wave channels. In the former, temperature effects are at most a few percent, while in the latter they are as large as 25 percent, for all temperatures under consideration. We stress that the temperature dependence reflects modifications in the spectrum and is not due to changes in susceptibilities or transport peaks. Since the latter appear around $\om\sim 0$, whereas NRQCD is an effective theory around the two-quark threshold or bound-state mass, $\om\sim 2m_b\sim M_{\rm bound}$,  these features are in fact not present in NRQCD.

\begin{figure}[t]
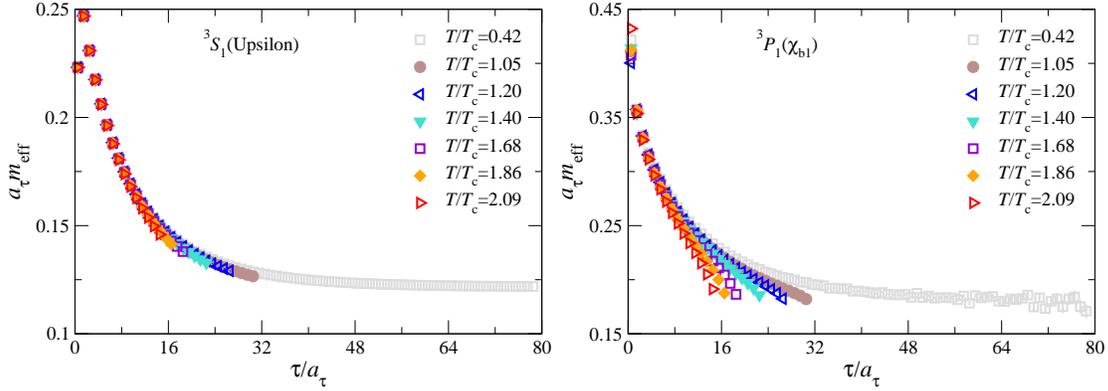

\begin{center}
\epsfig{figure=upsilon_all-T_eff-mass-GA.eps,width=0.48\textwidth}
\epsfig{figure=chi_b1_all-T_eff-mass-GA.eps,width=0.48\textwidth}
\end{center}
 \caption{Euclidean-time dependence of the effective mass, for different temperatures, in the $\Upsilon$ (left) and the $\chi_{b1}$ channel (right) \cite{Aarts:2010ek,Aarts:2013}.
}
 \label{fig:2}
\end{figure}

In Fig.\ \ref{fig:2} we show the effective masses, defined in the usual way,
\be
a_\tau m_{\rm eff} = -\log[G(\tau)/G(\tau-a_\tau)].
\ee
In the $S$ wave channel, we observe that the trend set by the lowest temperature is followed, even though there are thermal effects. In the $P$ wave channel on the other hand, we observe a deviation from the low-temperature trend, with the effective masses no longer showing an indication of reaching a plateau. We interpret this as a signal for melting, possibly immediate above $T_c$.

\section{Spectral functions}

To analyse in detail how the presence of the QGP affects the bottomonium system, we construct the spectral functions $\rho(\om)$ at each temperature. Since the inversion of Eq.\ (\ref{eq:1}) is an ill-posed problem---$G(\tau)$ contains only a finite number of data points, whereas $\rho(\om)$ is in principle a continuous function of $\om$---we use the Maximal Entropy Method (MEM) \cite{Asakawa:2000tr,Bryan} to achieve this. For a discussion of MEM and tests of the robustness of the results, we refer to our papers \cite{Aarts:2011sm,Aarts:2012ka,Aarts:2013}. We emphasise that it is essential to have very precisely determined correlators; the relative error in the $S$ wave channels is err$[G(\tau)]/G(\tau)\lesssim 4\times 10^{-4}$, for all euclidean times and temperatures. Also relevant is that due to the nonperiodicity of the correlators in NRQCD as well as the anisotropy, a large number of time slices is available for the MEM analysis, even at the highest temperature.

\begin{figure}[t]
\begin{center}
\epsfig{figure=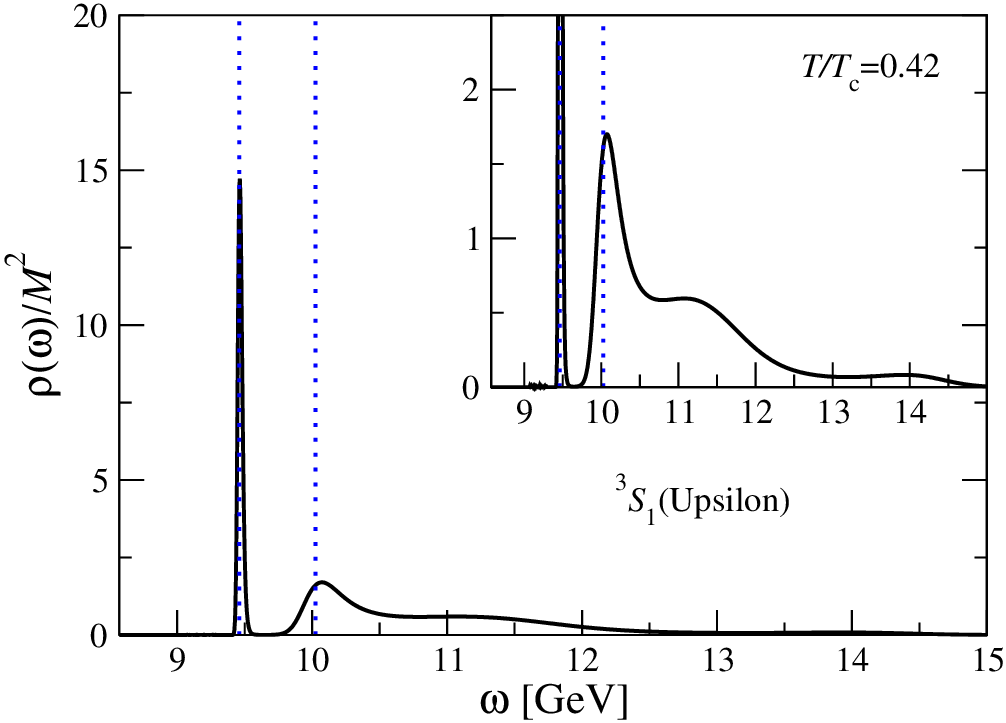,width=0.46\textwidth}
\epsfig{figure=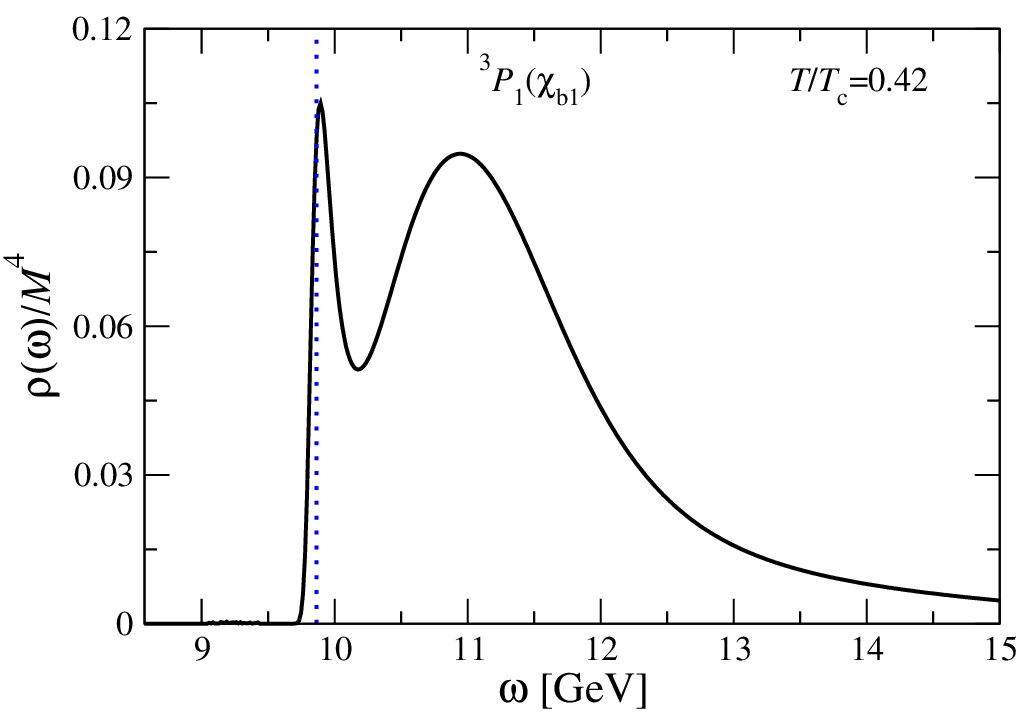,width=0.48\textwidth}
\end{center}
 \caption{Spectral function at the lowest temperature in the $\Upsilon$ channel \cite{Aarts:2011sm} (left) and the $\chi_{b1}$ channel \cite{Aarts:2013} (right). The dotted lines indicate the position of the ground state and first excited state (left only)  obtained with standard exponential fits.
 }
 \label{fig:T0}
\begin{center}
\epsfig{figure=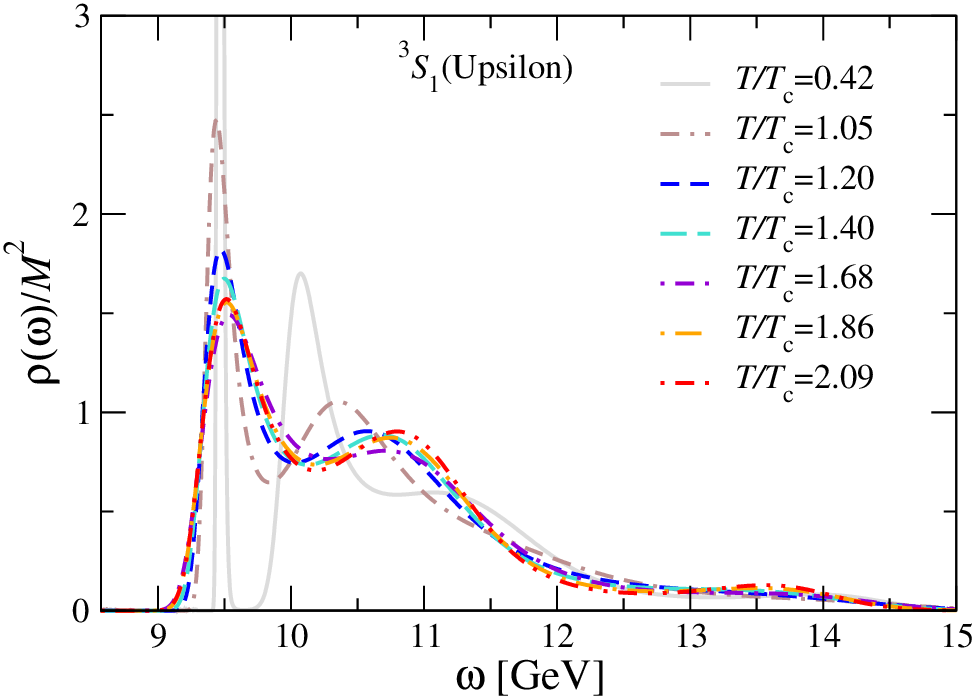,width=0.48\textwidth}
\epsfig{figure=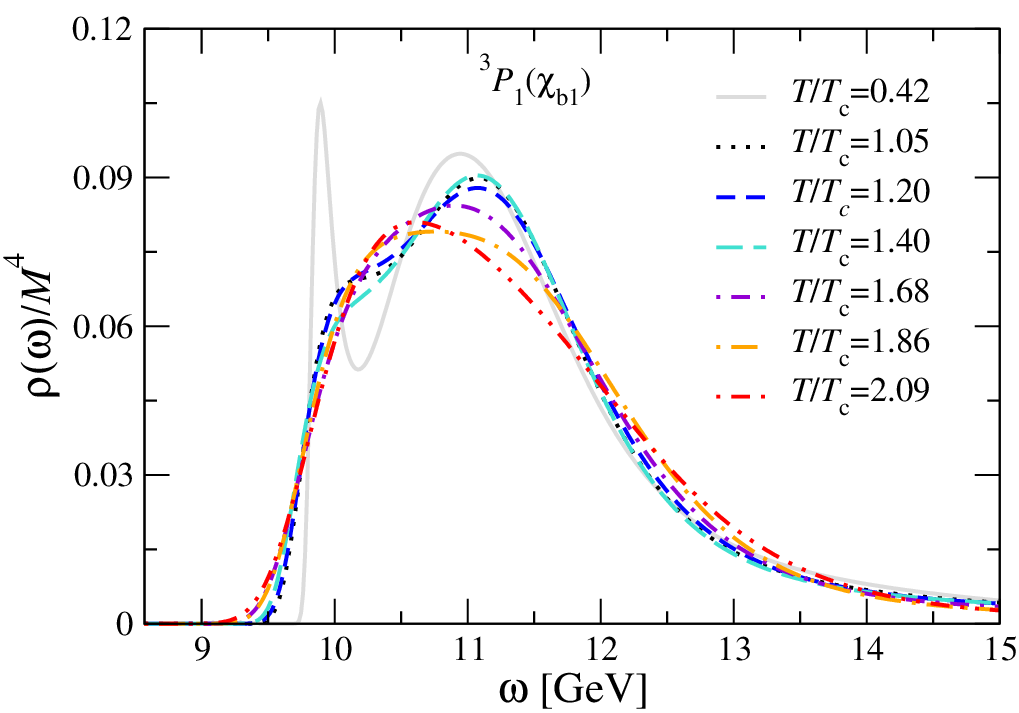,width=0.48\textwidth}
\end{center}
 \caption{As above, for all temperatures.}
 \label{fig:T}
\end{figure}

The spectral functions at the lowest temperature are shown in Fig.\ \ref{fig:T0}. The dotted lines indicate the position of the ground states and the first excited state (in the $\Upsilon$ channel only), obtained by standard exponential fits. We can therefore safely conclude that the first peak corresponds to the ground state and the second peak in the $\Upsilon$ channel to the first excited state. The structures at higher energies are presumably a mixture of further excited states and lattice artefacts, this will be discussed further below. It should be noted that in NRQCD the spectrum is determined up to one additive shift, which we determine from the $\Upsilon$ ground state. When converting energies in lattice units to GeV's, this implies that $a_\tau\om=0$ corresponds to $\om=8.57$ GeV in Figs.\ \ref{fig:T0} and \ref{fig:T}.

The results in the QGP are shown in Fig.\ \ref{fig:T}. In the $\Upsilon$ channel, the ground state remains visible, even though its width increases and its height decreases, i.e.\ we observe thermal broadening. Besides this, there is evidence for a shift of the peak position to the right, i.e.\ a thermal mass shift. Both phenomena are also present in EFT computations, relying on weak coupling \cite{Brambilla:2010vq}.
On the other hand, the peak corresponding to the first excited state below $T_c$ is no longer discernible above $T_c$. In an optimistic interpretation this can be seen  as a suppression of excited states, in line with results obtained at the LHC \cite{Chatrchyan:2011pe}.
These findings should be contrasted with the results for $P$ waves, demonstrated for the $\chi_{b1}$ channel in Fig.~\ref{fig:T} (right). Here we observe that above $T_c$ the ground state peak has disappeared from the spectrum. All that remains is a steeply rising spectral function.
Combined with the absence of indications for a plateau in the effective masses in the QGP, we are led to interpret this as a melting of  $P$ wave bottomonium states, already at $T=1.05T_c$.

\begin{figure}[t]
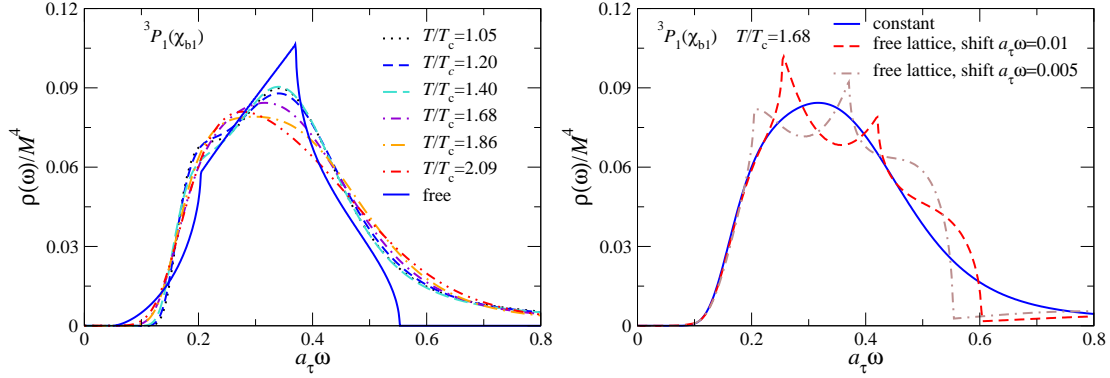

\begin{center}
\epsfig{figure=rho-chi_b1_all-aw-free2-GA.eps,width=0.48\textwidth}
\epsfig{figure=rho-chi_b1_free-recon-20-GA-v2.eps,width=0.48\textwidth}
\end{center}
 \caption{Left: Comparison between the spectral functions above $T_c$ in the $\chi_{b1}$ channel  (obtained with a constant default model) and the free lattice spectral function. Right: default model dependence, using a constant default model and the free lattice default model, shifted by $a_\tau\om=0.05$ and 0.1, at $T/T_c=1.68$ \cite{Aarts:2013}.}
 \label{fig:sys}
\end{figure}

If the $P$ wave bound states have melted, the spectral functions above $T_c$ should reflect quasi-free behaviour for the $b$ quarks in the QGP. 
In the absence of interactions, spectral functions in $P$ wave channels read, in the continuum \cite{Burnier:2007qm},
\be
\rho_P^{\rm free, cont}(\om) \sim \left(\om-\om_0\right)^{3/2}\Theta\left(\om-\om_0\right),
\ee
where $\om_0$ is the two-quark threshold.
However, on the lattice this simple expression is modified by lattice artefacts, due to the finiteness of the  Brillouin zone \cite{Aarts:2011sm}. In particular, the free lattice spectral function $\rho_P^{\rm free, lat}(\om)$ has support in a finite energy range only and the edges of the Brillouin zone show up as cusps. 
In Fig.\ \ref{fig:sys} (left) we show  $\rho_P^{\rm free, lat}(\om)$ together with the spectral functions above $T_c$.  Since the position of the threshold $\om_0$ depends on details of the lattice theory and the quark mass, we are allowed to shift $\rho_P^{\rm free, lat}(\om)$ horizontally. What cannot be changed, however, is the width of the interval in which the free spectral function has support and the relative positions of the cusps.
We see approximate agreement in the width of the region where the spectral weight is nonzero between the free and nonperturbatively determined spectral functions.
Therefore this lattice property indeed appears to be encoded in the euclidean correlators and the spectral functions obtained by MEM.

In the figures discussed above, we have used a constant default model in MEM. It is also possible to use $\rho_P^{\rm free, lat}(\om)$ as a default model directly. The resulting spectral functions are shown in Fig.\ \ref{fig:sys} (right) for $T/T_c=1.68$, using two different shifts of $\rho_P^{\rm free, lat}$. We observe that MEM is not capable of removing the nonanalytical behaviour of the cusps. Hence we favour smooth default models over nonanalytical ones. 
Nevertheless the onset of the spectral functions is in excellent agreement in all three default model cases. 
We also note that the region where lattice artefacts appear is uncomfortably close to the threshold, which indicates that a repetition of the analysis on lattices with a smaller spatial lattice spacing is certainly desirable.

\section{Outlook}

Our results indicate a melting pattern for bottomonium in which $S$ wave ground states survive but $P$ wave ground states melt at a temperature close to $T_c$. However, it should be kept in mind that all our conclusions are drawn from lattice ensembles at a single lattice spacing. In order to gain further confidence, it is therefore necessary to repeat the analysis on finer (spatial) lattices. We are currently extending our work to $N_f=2+1$ lattices to achieve precisely this and first results have already been reported at this meeting, not only for bottomonium \cite{tim}, but also for the electrical conductivity \cite{Amato:2013naa} and charge susceptibility \cite{Giudice:2013fza}.
Finally, it might also be worthwhile to use variants of MEM  \cite{Rothkopf:2011ef,Burnier:2013nla} to verify the robustness of the conclusions \cite{seyong}.

\acknowledgments

We thank Don Sinclair and Bugra Oktay for collaboration.
This work was partly supported by the European Community under the FP7 programme HadronPhysics3, 
 STFC, UKQCD and the STFC funded DiRAC Facility, 
 the Royal Society, the Wolfson Foundation, the Leverhulme Trust,
the National Research Foundation of Korea,  
Trinity Centre for High Performance Computing and the IITAC project funded by the HEA under the Program for Research in Third Level Institutes (PRTLI) co-funded by the Irish Government and the European Union, 
the Research Executive Agency (REA) of the European Union (ITN STRONGnet) and the Science Foundation Ireland.

\end{document}